\documentclass[conference]{IEEEtran}
\IEEEoverridecommandlockouts

\usepackage{tikz}
\usepackage{pgfplots}
\pgfplotsset{compat=1.17}
\usetikzlibrary{svg.path}
\usepackage{cite}
\usepackage{amsmath,amssymb,amsfonts}
\usepackage{graphicx}
\usepackage{textcomp}
\usepackage{xcolor}
\usepackage{hyperref}
\usepackage{cleveref}
\usepackage{commath}
\usepackage{algorithm}
\usepackage{algpseudocode}
\usepackage{optidef}
\usepackage{lipsum}
\usepackage{enumitem}
\usepackage{svg}
\usepackage{lettrine}
\usepackage{subcaption}
\usepackage{float}      
\usepackage{caption}    
\usepackage{array}      
\usepackage{booktabs}   
\usepackage[symbol]{footmisc}
\usepackage{geometry}
\usepackage{titlesec}   

\titlespacing*{\section}{0pt}{0.5\baselineskip}{0.3\baselineskip}
\titlespacing*{\subsection}{0pt}{0.4\baselineskip}{0.2\baselineskip}

\makeatletter
\newcommand{\nosemic}{\renewcommand{\@endalgocfline}{\relax}}
\newcommand{\dosemic}{\renewcommand{\@endalgocfline}{\algocf@endline}}

\let\oldnl\nl
\newcommand{\nonl}{\renewcommand{\nl}{\let\nl\oldnl}}
\makeatother

\renewcommand{\arraystretch}{1.25}

\def\BibTeX{{\rm B\kern-.05em{\sc i\kern-.025em b}\kern-.08em
    T\kern-.1667em\lower.7ex\hbox{E}\kern-.125emX}}

\usepackage{scalerel}
\definecolor{orcidlogocol}{HTML}{A6CE39}
\tikzset{
  orcidlogo/.pic={
    \fill[orcidlogocol] svg{M256,128c0,70.7-57.3,128-128,128C57.3,256,0,198.7,0,128C0,57.3,57.3,0,128,0C198.7,0,256,57.3,256,128z};
    \fill[white] svg{M86.3,186.2H70.9V79.1h15.4v48.4V186.2z}
                 svg{M108.9,79.1h41.6c39.6,0,57,28.3,57,53.6c0,27.5-21.5,53.6-56.8,53.6h-41.8V79.1z M124.3,172.4h24.5c34.9,0,42.9-26.5,42.9-39.7c0-21.5-13.7-39.7-43.7-39.7h-23.7V172.4z}
                 svg{M88.7,56.8c0,5.5-4.5,10.1-10.1,10.1c-5.6,0-10.1-4.6-10.1-10.1c0-5.6,4.5-10.1,10.1-10.1C84.2,46.7,88.7,51.3,88.7,56.8z};
  }
}

\newcommand\orcidicon[1]{\href{https://orcid.org/#1}{\mbox{\scalerel*{
\begin{tikzpicture}[yscale=-1,transform shape]
\pic{orcidlogo};
\end{tikzpicture}
}{|}}}}
\author{
  \IEEEauthorblockN{
    Ayesha Abid\IEEEauthorrefmark{1},
    Muhammad Jazib\IEEEauthorrefmark{1},
    Muhammad Riaz\IEEEauthorrefmark{2}
  }
  \IEEEauthorblockA{
    \IEEEauthorrefmark{1}\IEEEauthorrefmark{2}Dept. of Electrical Engineering, PIEAS, Pakistan
  }
  \IEEEauthorblockA{
    Email: \{bsee2068, bsee2023, riaz\}@pieas.edu.pk
  }
}

\begin{document}
\IEEEpubid{\makebox[\textwidth]{979-8-3315-1929-2/24/\$31.00~\copyright~2024 IEEE \hfill}}

\title{RCD-IoT:Enabling Industrial Monitoring and Control with Resource-Constrained Devices Under High Packet Transmission Rates}
\maketitle
\begin{abstract}
\let\thefootnote\relax\footnotetext{\IEEEauthorrefmark{1} The authors contributed equally and are co-first authors.}
 This paper highlights the significance of resource-constrained Internet of Things (RCD-IoT) systems in addressing the challenges faced by industries with limited resources. This paper presents an energy-efficient solution for industries to monitor and control their utilities remotely. Integrating intelligent sensors and IoT technologies, the proposed RCD-IoT system aims to revolutionize industrial monitoring and control processes, enabling efficient utilization of resources.The proposed system utilized the IEEE 802.15.4 WiFi Protocol for seamless data exchange between Sensor Nodes. This seamless exchange of information was analyzed through Packet Tracer. The system was equipped with a prototyped, depicting analytical chemical process to analyze the significant performance metrics. System achieved average Round trip time (RTT) of just 12ms outperforming the already existing solutions presented even with higher Quality of Service (QoS) under the transmission of 1500 packets/seconds under different line of sight (LOS) and Non line of sight (NLOS) fadings.
\end{abstract}


\section{Introduction}
\lettrine {I}{n} the era of Industry 4.0, the need for the industrial internet of things (IIoT) has become increasingly important as industries worldwide strive for enhanced efficiency, productivity,  energy and sustainability. The advancement is enabling real-time data collection, analysis, and decision-making. The companies can optimize processes and predict maintenance needs with interconnected devices and systems. The Industrial Internet of Things (IIoT) represents a significant advancement in the Industrial sector\cite{10339747}. This integration of information technology (IT) and electronics within the industry creates a smart environment and serves as a mechanism for monitoring and controlling diverse systems and devices. As automation steadily takes over traditional industrial methodologies, the global market for IIoT is projected to escalate to nearly USD 300 billion by 2024\cite{7004894}. 

Utilizing resource-constrained devices (RCD) in IoT for industrial monitoring and control offers many benefits. These devices are characterized by their efficiency in resource utilization. They ensure optimal performance within power, processing capabilities, and memory constraints. This efficiency reduces operational costs and energy consumption, making them particularly attractive for large-scale industrial deployments where cost-effectiveness is majorly concerned. Moreover, resource-constrained IoT devices are compact and have robust designs\cite{9328432}, which helps them achieve seamless integration into industries. 

\begin{figure}[htbp]
    \centering
    \includegraphics[width=1\columnwidth]{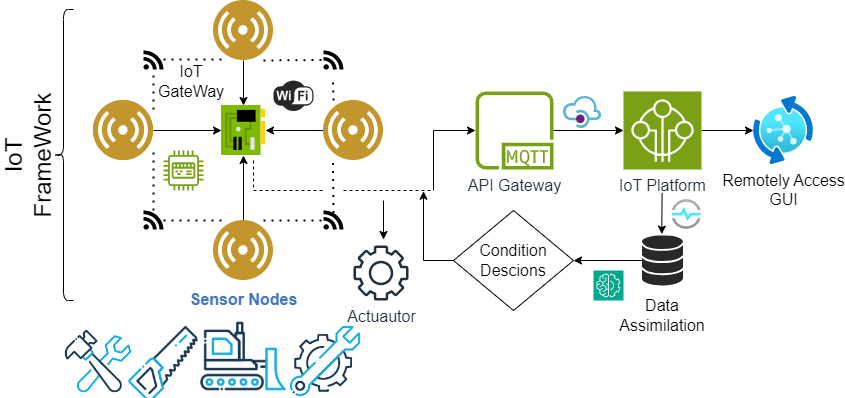} 
    \caption{Workflow Diagram of the Proposed Test Bed}
    \label{fig:1}
\end{figure}
In this paper, we propose an efficient and energy-optimized test bed for real-time industrial monitoring .\\
In \cite{8538692} author presented an latency evaluation of MQTT and Web socket protocol for industrial perspective.They have made a seamless communication between one server in Italy and one in Brazil, and acquired a almost a same average round-trip time of 150ms.
In \cite{s21175737} authors presented a comparison of QUIC and MQTT protocols and thier reliability, however their comparison under different fading scenarios were still lacking.
Similarly,\cite{8523891} the author gave an MQTT-based air quality monitoring system. Utilized a RCD esp8266 nodeMCU microcontroller that is connected to sensors to measure temperature, humidity, PM2.5, CO, and Ozone gas concentrations. Their research work also lacks an analysis of the system's performance in responsiveness and ability to maintain seamless responsiveness 
The existing MQTT based industrial micro controller lacks in terms of the latency of the system, which is considered a significant gap for the critical industrial infrastructures where a swift response is expected from the system.MQTT being its lighweight nature considered as an optimum communication protocol for industrial settings but still it is far behind in  repsonsevenes. In authors \cite{opavcin2023developing} have analyzed the response time of industrial microcontroller for three different industrial scenario however they have managed to least possible latency time was about 321.3ms for 1000messages/100ms.
 Authors \cite{inproceedings} demonstrated that Node-RED could be used to develop practical IoT applications quickly. At the same time, they showed that Node-RED has significant advantages in data visualization.Message queing telemetery transport prtocol was adopted in our work as it is considered as protocol with lightweight and low overhead makes it ideal for industrial settings where network resources in terms of bandwidth specifically in our case scenario where 1500 packets were transmitted oer 5 seconds with no packet loss\cite{8024687} are often limited.
 
\textbf{Contributions:}Inspired by the advantages of MQTT, our paper makes the following contributions to knowledge:
\begin{itemize}
    \item Presented a custom MQTT based solution for efficient communication in industrial setting, surpassing the efficacy of already proposed solution with an average round trip time of 12ms. 
    \item Implemented a real-time control mechanism based on MQTT which enhances system responsiveness. 
    \item Compared performance of our system under high QoS and packet transmission rates its ability to handle high traffic load makes it enable for larger wireless sensor networks considering thermal noises of  industrial utilities.
\end{itemize}
\section{IoT Protocols}
 Different IoT Protocols were compared\cite{9711544} before choosing the optimum one for our application; the comparative Analysis was done based on our requirements. We will discuss the most popular IoT protocols that are used for Internet of Things embedded systems and applications, along with their advantages and disadvantages and why we selected IEC 61850/IEC 60870-104 or Message Queuing Telemetry Transport (MQTT) protocol for our research application. It is essential to use a corrected IoT protocol that helps to decide the system's purpose, hardware, and software. There are so many IoT protocols, among which we will only cover Protocols - MQTT, CoAP, HTTP, and XMPP. 

\begin{table}[htbp]
\centering
\caption{Comparison of IoT Protocols}
\begin{tabular}{@{}l m{1.6cm} m{1.6cm} m{2.0cm}@{}}
\toprule
\textbf{Protocol} & \textbf{Description} & \textbf{Advantages} & \textbf{Disadvantages} \\
\midrule
MQTT & Lightweight, low power & Efficient & No security \\
CoAP & Web transfer for IoT & Low overhead & Small payload \\
HTTP & Standard for web & Ubiquitous & High overhead \\
XMPP & Real-time messaging & Extensible & Complex \\
\bottomrule
\end{tabular}
\label{tab:iot_comparison}
\end{table}

\textbf{MQTT-Protocol:}
A lightweight Communication Protocol has been utilized in our work to enhance the seamless communication between sensor nodes and the server. Before this, issues existed where packets/data loss occurred due to inappropriate communication protocols; MQTT utilizes the Publisher/Subscriber Model, and the QoS Concept ensures data transfer. We have used the appropriate QoS for our system design and analyzed the results by monitoring Traffic Analysis occurring at the same GateWay.The choice of this protocol was based on the power utilization\cite{8765692}
\begin{figure}[htbp]
  \centering
    \includegraphics[scale=0.35]{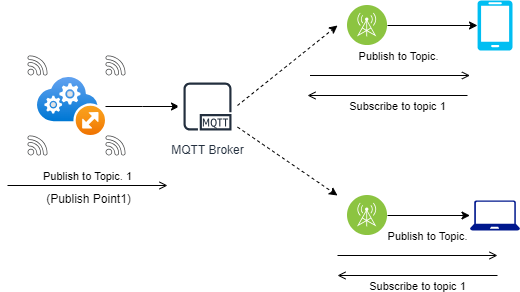}
    \caption{MQTT Publisher-Subscriber Model Workflow}
    \label{fig:6}
\end{figure}
\section{Working Methodology}
\subsection{Sensor Layer}
We developed a prototype setup resembling a conventional three-tank system in industrial processes. This setup aimed to emulate mixing procedures and facilitate chemical identification tasks. For experimentation, we used methyl orange in Tank 1 and sodium hydroxide (NaOH) in Tank 2. In the final monitoring tank, we observed the mixing process and chemical identification. The reaction between methyl orange and the base (NaOH) results in a color change to yellow, indicating a successful chemical interaction.
\begin{algorithm}
\caption{Communication Channel Simulation with Fading in Industrial IoT}
\begin{algorithmic}[1]
    \State \textbf{Input:} Number of sensor nodes ($N$), duration ($T$), fading type (\textit{fading})
    \State \textbf{Initialize:} $\mathbb{S} \gets$ $N$ sensor nodes with \textit{fading}, $\mathbb{C} \gets$ Central Server

    \For{$t = 1$ to $T$} 
        \State \textbf{Print} ``Time $t$''
        \For{each sensor $S_i \in \mathbb{S}$}
            \State \textit{packets\_sent} $\gets S_i.\textbf{send\_data}()$ with AWGN noise
            \State \textit{latency} $\gets S_i.\textbf{get\_latency}(QoS = 2)$
            \State \textbf{Apply Fading:}
            \If{\textit{fading} is Rayleigh} $\text{Fading} \gets \mathcal{R}(\sigma{=}1)$
            \ElsIf{\textit{fading} is Rician} $\text{Fading} \gets \text{Rician}(v{=}1.0, \sigma{=}1)$
            \ElsIf{\textit{fading} is AWGN} $\text{Fading} \gets \text{AWGN}(m{=}1.0, \omega{=}1.0)$
            \EndIf
            \State $\mathbb{C}.\textbf{receive\_data}($\textit{packets\_sent}, \textit{latency}$)$
        \EndFor
        \State \textit{average\_latency} $\gets \mathbb{C}.\textbf{get\_average\_latency}()$
        \State \textbf{Print} ``Central Server: Packets: $ \sum_{i=1}^N \textit{packets\_sent}_i$, Avg Latency: \textit{average\_latency} ms''
    \EndFor
    \State \textbf{Reset:} $\mathbb{C}.\textbf{reset}()$
\end{algorithmic}
\end{algorithm}

We carefully analyzed and selected efficient and cost-effective microcontrollers  for our prototype's core control and communication functionalities: the IEEE 802.11 Wi-Fi-supported device of ESP32 and the NodeMCU board of ESP8266. Both devices are widely used in industrial applications, and their compatibility with resource-constrained requirements makes them suitable for IoT projects. Their capabilities align well with the specific needs and constraints of our prototype.
\begin{figure}[t!]
  \centering
    \includegraphics[scale=0.45]{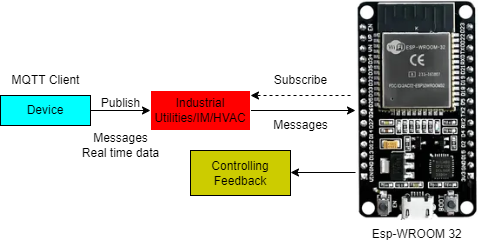}
    \caption{MQTT Communication Flow with ESP32}
    \label{fig:6}
\end{figure}
Incorporating sensors was essential to our hardware design, which enabled us to monitor and control functionalities precisely within the prototype setup. Specifically, we integrated the following sensors.
\begin{table}[htbp]
    \centering
    \caption{Comparison of Sensors and Applications}
    \label{tab:sensor_comparison}
    \renewcommand{\arraystretch}{1.3} 
    \resizebox{0.8\columnwidth}{!}{
        \begin{tabular}{|l|m{3.5cm}|} 
            \hline
            \textbf{Sensors} & \textbf{Application} \\
            \hline
            pH Sensor 4502-C & pH Measurement \\
            RGB TCS 230 & Colorimetric Sensing \\
            Ultrasonic HC-SR04 & Liquid Level Measurement \\
            DHT11 & Temperature and Humidity Sensing \\
            MQ-7 & Gas Concentration \\
            \hline
        \end{tabular}%
    }
\end{table}
The data collection process for the five sensors involves retrieving measurements from the physical environment and converting them into digital signals for processing utlizing Serial Perpiheral Interface (SPI) and Inter-integrated communication I2C protocol. Each sensor operates on a distinct working principle according to its application.

\subsection{Data Communication layer}
MQTT was utilized in our industrial setting. For its lightweight nature, efficient utilization of bandwidth MQTT protocol is at par.Latency calculation in MQTT based network setting takes place using Queueing theory and can be calculated using ersnal formula
\begin{equation}
W_q = \frac{\frac{\left( \frac{\lambda}{\mu} \right)^c \frac{c}{c!}}{1 - \frac{\lambda}{c \mu}}}
{\sum_{k=0}^{c-1} \frac{1}{k!} \left( \frac{\lambda}{\mu} \right)^k + \frac{\left( \frac{\lambda}{\mu} \right)^c}{c!} \frac{c \mu}{c \mu - \lambda}}
\end{equation}
Where \(\lambda\) is average arrival rate, \(\mu\) is service rate, \(c\) represents the number of servers, \(W_q\)average waiting time in the queue and \(\rho\) is traffic intensity, which is calculated as \(\rho = \frac{\lambda}{\mu}\) \cite{yokotani2021iot}. For stability, \(\rho < c\) must hold.\(k\)index used in the summation, which ranges from 0 to \(c-1\). 
MQTT provides QoS levels to ensure message deleivery reliability. Markov chain can model the state transitions for different QoS level\cite{chen2020study}. For making our system reliable at highest level. It ensures that message is delivered and acknowledged exaclty once between a sender and a reciever. Qos 2 can be modeled using this Markov chain transition matrix considering 5 states S1 idle S2 Publish(sent) S3(Pubrec recieved) S4 (Pubrel sent) S5(Pubcom recieved) with state vector as
\begin{equation}
\pi = [\pi_1, \pi_2, \pi_3, \pi_4, \pi_5]
\end{equation}
The transition matrix P can be defined as follows 
\[
P = \begin{bmatrix}
1 - \alpha & 0 & 0 & 0 & \delta \\
\alpha & 1 - \beta & 0 & 0 & 0 \\
0 & \beta & 1 - \gamma & 0 & 0 \\
0 & 0 & \gamma & 1 - \delta & 0 \\
0 & 0 & 0 & \delta & 0
\end{bmatrix}
\]
Each diagonal element represents the probability of remaining in the same rate due to failure at that stage. The reliability R of QoS 2 can be calculated by the following mathematical equation as the probability of reaching state S5 successful delivery of the message 
\[
R = \pi_5 = \frac{\alpha \beta \gamma \delta}{(1 - (1 - \alpha) (1 - \beta) (1 - \gamma) (1 - \delta))}
\]

Where:

\begin{itemize}
    \item $R$ represents the overall reliability of delivering the message exactly once.
    \item The product $\alpha \beta \gamma \delta$ represents the probability that all steps are completed successfully without any failure.
\end{itemize}

\subsection{Network Analysis}
Round Trip Time with Multi hop  is a significant meteric to eavlaute communication latency in IoT Applications\cite{8078359}
The RTT for the MQTT protocol can be calculated as follows:
\begin{equation}
RTT = 2 \sum_{i=1}^{n} \left( \frac{d_i}{v_i} + \frac{L_i}{R_i} h_i + P_i + \frac{h_i}{\mu_i - \lambda_i} \right) + \sum_{i=1}^{n} \frac{T_{\text{transmission}_i}}{1 - p_i}
\end{equation}

where n represents the number of hops between the source and the destination in the network. The variable \(d_i\) is the distance between nodes at the \(i\)th hop (in meters), while \(v_i\) represents the propagation speed of the signal at the \(i\)th hop. The length of the packet transmitted at the \(i\)-th hop is denoted by \(L_i\) (in bits), and \(R_i\) is the transmission rate at the \(i\)-th hop.

The processing delay at the \(i\)-th hop, represented by \(P_i\), indicates the time required for the microcontroller to process the packet\cite{zhu2013quantitative}. The average arrival rate of packets at the \(i\)th hop is denoted by \(\lambda_i\) (in packets per second), while \(\mu_i\) is the average service rate at the \(i\)th hop. 
\subsection{Communication Channel}
To mimic the real world industrial communication scenario where the thermal noise of industrial utilities could fade the transmitter signal in wireless commuincation channel hence included the AWGN noise , rayleigh and ricieian fading in our wireless channel.
\textbf{AWGN (Additive White Gaussian Noise) Model:} The AWGN channel is a basic model used to represent the effect of noise on a signal passing through a medium. This noise is modeled as Gaussian-distributed with a constant spectral density. The mathematical representation for the AWGN model is given by:
\begin{equation}
y(t) = x(t) + n(t)
\end{equation}
where \( y(t) \in \mathbb{C} \) is the received complex signal, \( x(t) \in \mathbb{C} \) is the transmitted complex signal, and \( n(t) \in \mathbb{C} \) is the noise term modeled as a complex Gaussian random variable:
\begin{equation}
n(t) \sim \mathcal{CN}(0, N_0)
\end{equation}
The probability density function (PDF) for the AWGN noise in the complex domain is given by:
\begin{equation}
f_{n}(n) = \frac{1}{\pi N_0} \exp\left( -\frac{|n|^2}{N_0} \right)
\end{equation}
\textbf{Rayleigh Fading Model:}
The Rayleigh fading channel is used to model a scenario where there are multiple paths for a signal to travel, with no direct Line of sight (NLOS)\cite{saadaoui2017iwsn}. The received signal under Rayleigh fading is given by:
\begin{equation}
y(t) = h(t) x(t) + n(t)
\end{equation}
where \( h(t) \in \mathbb{C} \) is the complex channel gain modeled as a Rayleigh random variable, and \( n(t) \) is the AWGN noise term. In general, the channel gain \( h(t) \) can be represented as:
\begin{equation}
h(t) = h_I(t) + j h_Q(t)
\end{equation}
where \( h_I(t) \) and \( h_Q(t) \) are independent, identically distributed (i.i.d.) Gaussian random variables with zero mean and variance \(\sigma^2\):
\begin{equation}
h_I(t), h_Q(t) \sim \mathcal{N}(0, \sigma^2)
\end{equation}
The magnitude \(|h(t)|\) follows a Rayleigh distribution with the PDF:
\begin{equation}
f_{|h|}(r) = \frac{r}{\sigma^2} \exp\left( -\frac{r^2}{2\sigma^2} \right), \quad r \geq 0
\end{equation}
The autocorrelation function of the fading channel, \( R_h(\tau) \), can be used to derive the Power Spectral Density (PSD). Assuming a Jakes' spectrum model, the PSD is given by:
\begin{equation}
S_h(f) = \frac{1}{\pi f_D \sqrt{1 - (f/f_D)^2}}, \quad |f| \leq f_D
\end{equation}
where \( f_D \) is the maximum doppler shift.

\textbf{Rician Fading Model:}
For environments with a dominant LOS path in addition to multiple scattered paths, the Rician fading channel is used\cite{liu2022tacan}. The received signal model is given by:
\begin{equation}
y(t) = h(t) x(t) + n(t)
\end{equation}
where the channel gain \( h(t) \) is given by:
\begin{equation}
h(t) = h_\text{LOS} + h_\text{NLOS}
\end{equation}
\begin{equation}
h_\text{LOS} = A e^{j \theta}
\end{equation}
where \( A \) is the amplitude of the LOS path and \( \theta \) is the phase. The Non Line of Sight (NLOS) component \( h_\text{NLOS} \) is modeled as a complex Gaussian random variable:
\begin{equation}
h_\text{NLOS} \sim \mathcal{CN}(0, 2\sigma^2)
\end{equation}
The magnitude \(|h(t)|\) of the Rician fading channel follows a Rician distribution with the PDF:
\begin{equation}
f_{|h|}(r) = \frac{r}{\sigma^2} \exp\left( -\frac{r^2 + A^2}{2\sigma^2} \right) I_0\left( \frac{r A}{\sigma^2} \right), \quad r \geq 0
\end{equation}
where \( I_0(\cdot) \) is the modified Bessel function of the first kind and zero-order. The Rician \( K \)-factor, which represents the ratio of the power in the direct LOS component to the scattered power, is given by:
\begin{equation}
K = \frac{|h_\text{LOS}|^2}{2\sigma^2} = \frac{A^2}{2\sigma^2}
\end{equation}
The autocorrelation of the Rician fading channel is given by:
\begin{equation}
R_h(\tau) = A^2 + 2\sigma^2 J_0(2\pi f_D \tau)
\end{equation}
where \( J_0(\cdot) \) is the Bessel function of the first kind and zero-order, and \( f_D \) is the maximum doppler shift.
All these commuincation scenarios are modeled on the basis of the industrial environment so that it could mimic real world commuincation scenario.
\section{Key Performance Metrics}
\subsection{Round Trip Time}
After deploying our essential hardware and software, we analyzed our network latency's Round-Trip Times (RTT). We deduced significant results regarding RTT, which are shown in the table.
\begin{table}[htbp]
    \centering
    \scriptsize 
    \caption{Summary of Round Trip Times (RTT)}
    \label{tab:RTT_summary}
    \renewcommand{\arraystretch}{1.2} 
    \resizebox{0.8\columnwidth}{!}{
        \begin{tabular}{|l|c|}
            \hline
            \textbf{Round Trip Time} & \textbf{Time in Milli Seconds} \\
            \hline
            Minimum RTT & 3 ms \\
            Maximum RTT & 72 ms \\
            Average RTT & 12 ms \\
            \hline
        \end{tabular}%
    }
\end{table}
Our results observed an average of 12ms of RTT, which shows how effective our MQTT Protocol was in the provided network.
\begin{figure}[htbp]
    \centering
    \includegraphics[width=0.3\textwidth]{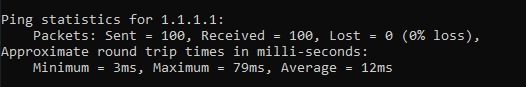}
    \caption{RTT Measurements Over Time}
    \label{fig:your_figure_label}
\end{figure}

\subsection{System Software}
Node-RED is a flow-based development tool for visual programming. Data from sensors interfaced with an ESP32 micro-controller is sent to Node-RED using the MQTT protocol Publisher/Subscriber Model as follows: The ESP32, acting as an MQTT client, gathers sensor data and publishes it to specified topics on an MQTT broker. The MQTT broker we have hosted locally manages the communication and ensures that data is distributed to the appropriate subscribers. Node-RED is configured as an MQTT client and subscribes to the relevant topics on the Public MQTT broker. When the ESP32 publishes sensor data, the MQTT broker forwards these messages to Node-RED, which processes, visualizes, or triggers workflows based on the received data, enabling seamless integration and real-time data handling.
\subsection{Adapter for Loop back Traffic}
\begin{figure}[htbp]
    \centering
\includegraphics[width=0.45\textwidth]{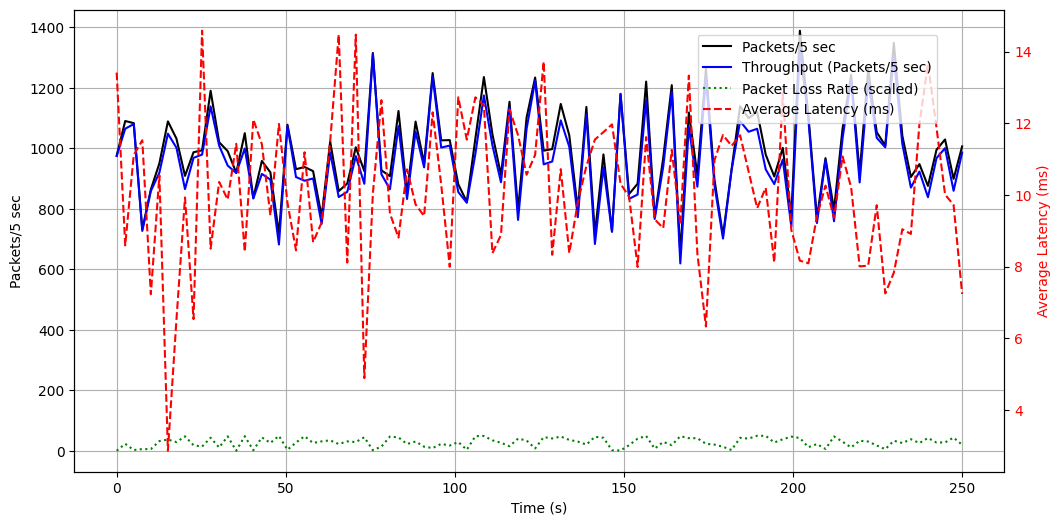}
    \caption{Packet Transmission Rate Over 5-Second Intervals}
    \label{alt}
\end{figure}

Our analysis of network capacity indicates that the network can accommodate the packet transmission rate without experiencing congestion or bandwidth limitations; the findings of this study reveals that sending 1500 packets / 5 seconds is feasible under certain conditions, provided that the network infrastructure, device capabilities, application requirements, QoS mechanisms, and real-time constraints are adequately addressed. These performance metrics were observed with zero packet loss.The results obtained can bee seen in Fig,~\ref{alt}.
\begin{figure}[htbp]
    \centering
\includegraphics[width=0.5\textwidth]{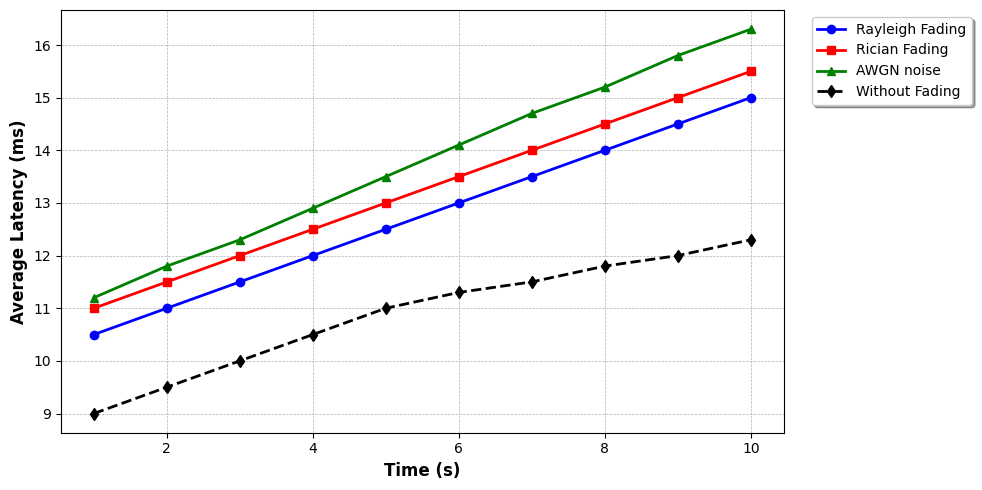}
    \caption{Comparison of faded vs non faded latencies}
    \label{faded vs non faded}
\end{figure}
Fig.~\ref{faded vs non faded} shows with increase in fading the wireless channel link gets weaker the and average latency increased with that.

\subsection{Graphical User Interface}
In our Node-RED flow, we configured nodes to collect data from sensors. Once the data is processed and ready for visualization, we set up nodes in Node-RED to transmit this data to the HINET Live Panel platform. This transmission typically involves utilizing the MQTT communication protocol supported by the HINET Live Panel.

Notably, the HINET Live Panel is connected to Node-RED, where it subscribes to specific topics. These topics are linked to the I/O variables created in our system, ensuring that the data is seamlessly received and displayed on the HINET Live Panel interface. On the HINET Live Panel platform, we have developed a GUI (Graphical User Interface) to visualize the incoming data in a user-friendly format. 

With the data flowing from Node-RED to the HINET Live Panel, users can monitor and analyze it in real-time through the GUI we created. They can view trends, anomalies, and insights from the data, enabling them to make informed decisions or take appropriate actions.

Depending on our requirements, we can further customize the GUI in the HINET Live Panel to add new features, enhance existing visualizations, or integrate additional data sources. We developed a control panel within the HINET Live Panel to enable the real-time operation of water pumps or motors. With this feature, users can remotely turn pumps on or off, providing greater control and efficiency in managing industrial processes. Furthermore, ongoing refinement of our Node-RED flows can optimize data processing and enhance integration with the HINET Live Panel, ensuring seamless operation and maximum utility of our monitoring and control system.

We have integrated alarm-triggering capabilities into our Node-RED flow and incorporated alarm notifications into the HINET Live Panel interface. This system provides real-time data visualization and enables proactive responses to critical events by triggering alarms when values exceed setpoints or thresholds. This feature ensures timely attention to potential issues, enhancing operational efficiency and reliability.
\section{Results}
In this section, we presented the experimental results on the impact of various fading and noise conditions, including Rayleigh fading, Rician fading, and AWGN, on industrial IoT network performance simulated and practically done. The metrics analyzed include average latency, packet transmission rates, throughput, and packet loss.
\begin{figure}[htbp]
    \centering
\includegraphics[width=0.3\textwidth]{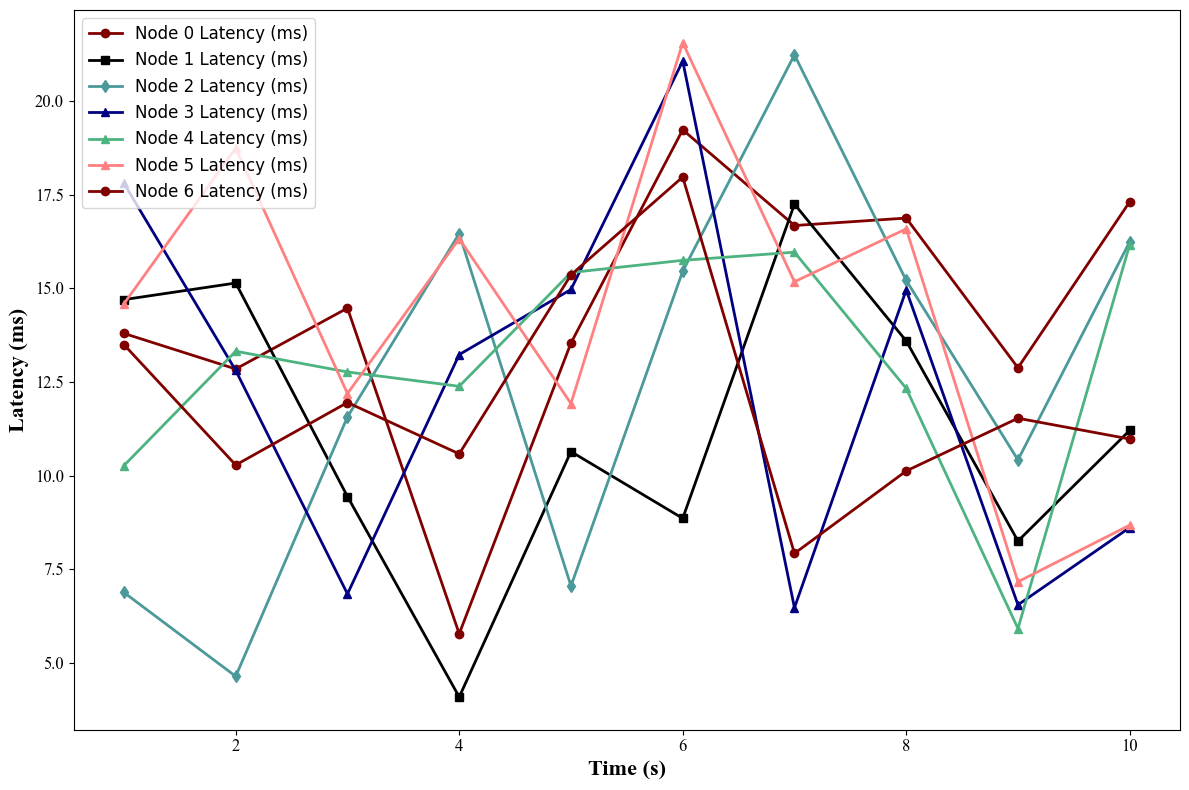}
    \caption{Average latency of all sensor nodes}
    \label{avg latency all }
\end{figure}
Table~\ref{tab:average_latency} summarizes the average latency for different channel conditions. AWGN noise resulted in the highest latency, increasing from 12.5 ms at 2 seconds to 16.0 ms at 10 seconds, indicating significant challenges in environments with high noise levels. Rayleigh and Rician fading conditions exhibited lower latency, with Rician fading outperforming Rayleigh due to the presence of a LOS component, which mitigates the effects of multipath fading. The lowest latency was observed in the no-fading scenario, demonstrating that minimizing fading is essential for achieving optimal performance in real-time applications.Fig.~\ref{alt}illustrates the packet transmission rate, throughput, packet loss rate, and average latency over time. Packet transmission rates fluctuated between 800 and 1200 packets per 5 seconds, with corresponding variations in throughput. Latency fluctuated between 8 ms and 14 ms, with peaks aligning with drops in throughput, suggesting temporary congestion or fading. Despite minor spikes in packet loss, the overall packet loss rate remained low, demonstrating effective packet handling by the network.

Fig.~\ref{faded vs non faded} provides a comparison of average latency across different conditions. The latency under AWGN noise was consistently the highest, while the no-fading scenario had the lowest latency. Rician fading demonstrated better performance compared to Rayleigh due to its LOS path, resulting in greater stability. The increasing trend in latency for all conditions over time underscores the need for robust congestion control and quality of service (QoS) mechanisms to maintain stable network performance.
\begin{table}[t!]
    \centering
    \caption{Average Latency for Various Fading and Noise Conditions in Industrial IoT Environments}
    \label{tab:average_latency}
    \renewcommand{\arraystretch}{1.8}
    \resizebox{\columnwidth}{!}{%
        \begin{tabular}{|c|c|c|c|c|}
            \hline
            & \textbf{Rayleigh Fading (ms)} & \textbf{Rician Fading (ms)} & \textbf{AWGN Noise (ms)} & \textbf{Without Fading (ms)} \\
            \hline
            \hline
            \textbf{Time = 2 s} & 11.0 & 11.5 & 12.5 & 9.5 \\
            \textbf{Time = 4 s} & 11.8 & 12.3 & 13.4 & 10.3 \\
            \textbf{Time = 6 s} & 12.5 & 13.0 & 14.2 & 11.0 \\
            \textbf{Time = 8 s} & 13.2 & 13.8 & 15.1 & 11.7 \\
            \textbf{Time = 10 s} & 14.0 & 14.5 & 16.0 & 12.5 \\
            \hline
        \end{tabular}%
    }
\end{table}
The result shows that under effective industrial environment our comparitive analysis sets a bench mark to make the latency loop holes better in industrial IoT environments. 
\section*{Conclusion and Future Work}
We have Presented an energy effecient solution for monitoring and controlling industrial Process at high packets transmission rates utilizing MQTT protocol with minimal average latency with established comparative analysis.All prototype industrial utilities were monitored and on the basis of monitoring it was controlled efficiently with an average RTT time of 12ms which states that our system was responsive enough to handle complex tasks.Our  system is just a prototype to analyze industrial environments with 5 sensors and some controlling mechanism to analyze the MQTT Protocol.This system architecture can be deployed and analyzed for industries other than Liquid based Application with some modification of sensors or hardware Furthermore, Predictive Maintenance of these smart sensors can also be done by obtaining faulty data of these sensors and classify them accordingly also a Digital Twin can be created for these sensors to handle data sparsity if there would be.
\bibliographystyle{IEEEtran}
\bibliography{ref}

\begin{thebibliography}{10}
\providecommand{\url}[1]{#1}
\csname url@samestyle\endcsname
\providecommand{\newblock}{\relax}
\providecommand{\bibinfo}[2]{#2}
\providecommand{\BIBentrySTDinterwordspacing}{\spaceskip=0pt\relax}
\providecommand{\BIBentryALTinterwordstretchfactor}{4}
\providecommand{\BIBentryALTinterwordspacing}{\spaceskip=\fontdimen2\font plus
\BIBentryALTinterwordstretchfactor\fontdimen3\font minus \fontdimen4\font\relax}
\providecommand{\BIBforeignlanguage}[2]{{%
\expandafter\ifx\csname l@#1\endcsname\relax
\typeout{** WARNING: IEEEtran.bst: No hyphenation pattern has been}%
\typeout{** loaded for the language `#1'. Using the pattern for}%
\typeout{** the default language instead.}%
\else
\language=\csname l@#1\endcsname
\fi
#2}}
\providecommand{\BIBdecl}{\relax}
\BIBdecl

\bibitem{10339747}
A.~Bashir, M.~A. Mohsin, M.~Jazib, and H.~Iqbal, ``Mindtwin ai: Multiphysics informed digital-twin for fault localization in induction motor using ai,'' in \emph{2023 International Conference on Big Data, Knowledge and Control Systems Engineering (BdKCSE)}, 2023, pp. 1--8.

\bibitem{7004894}
C.~Perera, C.~H. Liu, S.~Jayawardena, and M.~Chen, ``A survey on internet of things from industrial market perspective,'' \emph{IEEE Access}, vol.~2, pp. 1660--1679, 2014.

\bibitem{9328432}
V.~A. Thakor, M.~A. Razzaque, and M.~R.~A. Khandaker, ``Lightweight cryptography algorithms for resource-constrained iot devices: A review, comparison and research opportunities,'' \emph{IEEE Access}, vol.~9, pp. 28\,177--28\,193, 2021.

\bibitem{8538692}
D.~R.~C. Silva, G.~M.~B. Oliveira, I.~Silva, P.~Ferrari, and E.~Sisinni, ``Latency evaluation for mqtt and websocket protocols: an industry 4.0 perspective,'' in \emph{2018 IEEE Symposium on Computers and Communications (ISCC)}, 2018, pp. 01\,233--01\,238.

\bibitem{s21175737}
\BIBentryALTinterwordspacing
F.~Fernández, M.~Zverev, P.~Garrido, J.~R. Juárez, J.~Bilbao, and R.~Agüero, ``Even lower latency in iiot: Evaluation of quic in industrial iot scenarios,'' \emph{Sensors}, vol.~21, no.~17, 2021. [Online]. Available: \url{https://www.mdpi.com/1424-8220/21/17/5737}
\BIBentrySTDinterwordspacing

\bibitem{8523891}
S.~Chanthakit and C.~Rattanapoka, ``Mqtt based air quality monitoring system using node mcu and node-red,'' in \emph{2018 Seventh ICT International Student Project Conference (ICT-ISPC)}, 2018, pp. 1--5.

\bibitem{opavcin2023developing}
S.~Opa{\v{c}}in, L.~Rizvanovi{\'c}, B.~Leander, S.~Mubeen, and A.~{\v{C}}au{\v{s}}evi{\'c}, ``Developing and evaluating mqtt connectivity for an industrial controller,'' in \emph{2023 12th Mediterranean Conference on Embedded Computing (MECO)}.\hskip 1em plus 0.5em minus 0.4em\relax IEEE, 2023, pp. 1--5.

\bibitem{inproceedings}
K.~Ferencz and J.~Domokos, ``Using node-red platform in an industrial environment,'' 02 2020.

\bibitem{8024687}
S.~Katsikeas, K.~Fysarakis, A.~Miaoudakis, A.~Van~Bemten, I.~Askoxylakis, I.~Papaefstathiou, and A.~Plemenos, ``Lightweight secure industrial iot communications via the mq telemetry transport protocol,'' in \emph{2017 IEEE Symposium on Computers and Communications (ISCC)}, 2017, pp. 1193--1200.

\bibitem{9711544}
T.~Anitha, S.~Manimurugan, S.~Sridhar, S.~Mathupriya, and G.~C.~P. Latha, ``A review on communication protocols of industrial internet of things,'' in \emph{2022 2nd International Conference on Computing and Information Technology (ICCIT)}, 2022, pp. 418--423.

\bibitem{8765692}
J.~Toldinas, B.~Lozinskis, E.~Baranauskas, and A.~Dobrovolskis, ``Mqtt quality of service versus energy consumption,'' in \emph{2019 23rd International Conference Electronics}, 2019, pp. 1--4.

\bibitem{yokotani2021iot}
T.~Yokotani, S.~Ohno, H.~Mukai, and K.~Ishibashi, ``Iot platform with distributed brokers on mqtt,'' \emph{International Journal of Future Computer and Communication}, vol.~10, no.~1, pp. 7--12, 2021.

\bibitem{chen2020study}
F.~Chen, Y.~Huo, K.~Liu, W.~Tang, J.~Zhu, and Z.~Sui, ``A study on mqtt node selection,'' in \emph{2020 16th International Conference on Mobility, Sensing and Networking (MSN)}.\hskip 1em plus 0.5em minus 0.4em\relax IEEE, 2020, pp. 622--623.

\bibitem{8078359}
P.~Ferrari, E.~Sisinni, D.~Brandão, and M.~Rocha, ``Evaluation of communication latency in industrial iot applications,'' in \emph{2017 IEEE International Workshop on Measurement and Networking (M\&N)}, 2017, pp. 1--6.

\bibitem{zhu2013quantitative}
N.~Zhu, J.~He, Y.~Zhou, and W.~Wang, ``Quantitative analysis of the correlation of round-trip times between network nodes,'' \emph{International Journal of Future Generation Communication and Networking}, vol.~6, no. 2ui, 2013.

\bibitem{saadaoui2017iwsn}
S.~Saadaoui, M.~Tabaa, F.~Monteiro, M.~Chehaitly, A.~Dandache, and A.~Oukaira, ``Iwsn under an industrial wireless channel in the context of industry 4.0,'' in \emph{2017 29th International Conference on Microelectronics (ICM)}.\hskip 1em plus 0.5em minus 0.4em\relax IEEE, 2017, pp. 1--4.

\bibitem{liu2022tacan}
F.~Liu, X.~Dai, M.~Jin, W.~Zhang, Y.~Yang, and F.~Qin, ``Tacan: The shaping of delay distribution under multipath fading channel for industrial iot systems,'' \emph{IEEE Internet of Things Journal}, vol.~9, no.~17, pp. 16\,714--16\,725, 2022.

\end{thebibliography}

\end{document}